\documentclass[aps,prd,secnumarabic,amssymb, amsmath,nobibnotes,nofootinbib,11pt]{revtex4}
\usepackage{amsfonts,amsmath,hyperref,url, color}
\usepackage{bm, bbm}

\newcommand{\be}{\begin{equation}}
\newcommand{\ee}{\end{equation}}
\newcommand{\bea}{\begin{eqnarray}}
\newcommand{\eea}{\end{eqnarray}}

\usepackage{color}

\begin{document}

\title{Deformed discrete symmetries}
\author{Michele Arzano}
\email{michele.arzano@roma1.infn.it}
\affiliation{Dipartimento di Fisica and INFN,\\ `Sapienza' University of Rome,\\ P.le A. Moro 2, 00185 Roma, EU }
\author{Jerzy Kowalski-Glikman}
\email{jerzy.kowalski-glikman@ift.uni.wroc.pl}
\affiliation{Institute for Theoretical Physics, University of Wroc\l{}aw, Pl.\ Maxa Borna 9, Pl--50-204 Wroc\l{}aw, Poland.}

\begin{abstract}
\begin{center}
{\bf Abstract}
\end{center}
We construct discrete symmetry transformations for deformed relativistic kinematics based on group valued momenta. We focus on the specific example of $\kappa$-deformations of the Poincar\'e algebra with associated momenta living on (a sub-manifold of) de Sitter space. Our approach relies on the description of quantum states constructed from deformed kinematics and the observable charges associated with them. The results we present provide the first step towards the analysis of  experimental bounds on the deformation parameter $\kappa$ to be derived via precision measurements of discrete symmetries and CPT.
\end{abstract}

\maketitle

Discrete symmetries play a fundamental role in our understanding of the microscopic constituent of matter and their interactions. The CPT theorem is one of the pillars of local quantum field theory and tests of CPT invariance can be performed with extraordinary sensitivity using neutral kaon systems \cite{Babusci:2013gda,Ambrosino:2006vr,DiDomenico:2010zz}. Such sensitivity can be used to put stringent constraints on scenarios in which the postulates at the basis of the CPT theorem are altered like, for example, in models featuring Lorentz symmetry violation. It is thus quite natural to ask what is the fate of CPT symmetries in scenarios in which the structure of the Poincar\'e group, instead, is altered in a way which renders possible the introduction of an invariant Planckian energy scale. Such models have been suggested as a `flat-spacetime limit' \cite{AmelinoCamelia:2003xp}, or the ``relative locality limit'' of quantum gravity \cite{AmelinoCamelia:2011bm}, describing Planck-scale deformations of relativistic kinematics. Even though their direct connection with quantum gravity has not yet been decisively proved, these models have played a prominent role as a test-bed for quantum gravity phenomenology. Most of such phenomenology has focused on the role of Planck-scale deformations of the energy-momentum dispersion relation and its impact on the time-of-flight of high energy particles from astrophysical sources (see e.g. \cite{AmelinoCamelia:2008qg}). In this note we discuss how discrete symmetries can be defined in a specific class of models of deformed symmetries based on non-abelian momentum group manifold. Our analysis should be seen as a first theoretical step towards the use of high sensitivity tests of CPT invariance in neutral kaon systems to set bounds on the energy scale characterizing the curvature of momentum space.

The key property of models of deformed relativistic symmetries we will consider is that ordinary flat Minkowski momentum space is replaced by a {\it curved, non-abelian group manifold} \cite{KowalskiGlikman:2003we}. In particular for the so-called $\kappa$-deformed Poincar\'e symmetries \cite{Lukierski:1991pn, Lukierski:1992dt, Majid:1994cy}, the momentum space is the $AN(3)$ group, a subgroup of the four-dimensional de Sitter (or the five-dimensional Lorentz) group $SO(4,1)$. Geometrically such space describes a submanifold of four dimensional de Sitter space. If one considers four-dimensional de Sitter space of radius $\kappa$, being a (hyper)-surface in the five dimensional Minkowski space defined by
\begin{equation}\label{1}
   -P_0^2 + P_1^2 + P_2^2 + P_3^2 + P_4^2 =\kappa^2\, ,
\end{equation}
the $AN(3)$ manifold is the submanifold of (\ref{1}) identified by the inequality
\begin{equation}\label{2}
   P_0 + P_4
    >0\, .
\end{equation}
Here we assume that $P_4$ is given by the positive root of
\begin{equation}\label{2a}
   P_4
    =\sqrt{\kappa^2 +P_0^2-\mathbf{P}^2}\doteq\sqrt{\kappa^2 +M^2}\, ,
\end{equation}
where $\doteq$ means equality on-shell, $P_0^2-\mathbf{P}^2=M^2$.

%There are, of course, many coordinates that cover the $AN(3)$ group manifold, (\ref{1}), (\ref{2}). The one we will be using quite often (in exchange of the coordinates $P_A$ defined above) are defined by
%\begin{eqnarray}
% {P_0}(k_0, \mathbf{k}) &=&  \kappa\sinh
%{\frac{k_0}\kappa} + \frac{\mathbf{k}^2}{2\kappa}\,
%e^{  {k_0}/\kappa}, \nonumber\\
% P_i(k_0, \mathbf{k}) &=&   k_i \, e^{  {k_0}/\kappa}, \nonumber\\
% {P_4}(k_0, \mathbf{k}) &=& \kappa \cosh
%{\frac{k_0}\kappa} - \frac{\mathbf{k}^2}{2\kappa}\, e^{  {k_0}/\kappa}.
%\label{3}
%\end{eqnarray}
%It is straightforward to check that the conditions (\ref{1}), (\ref{2}) are satisfied.

There is a natural action of the Lorentz group $SO(3,1)$ on the manifold (\ref{1}) (it just leaves $P_4$ invariant and $P_\mu$ transform as components of a four-vector). The orbits of $SO(3,1)$ on the $AN(3)$ group will span positive and negative energy mass-shells. Notice however that the condition (\ref{2}) is clearly not Lorentz invariant, for the standard Lorentz action. A solution to this problem was suggested in \cite{Arzano:2009ci}, where it was pointed out that the Lorentz action on negative energy states has to be modified to involve the notion of  `antipode'. Here we would like to point out that such modification should not be seen as an `ad hoc' prescription. Indeed, as in usual QFT, states with support on the negative mass-shell are elements of the complex conjugate one-particle Hilbert space, which is naturally isomorphic to the {\it dual} Hilbert space \cite{Geroch:1985ci}. Thus the action of symmetry generators on states with support on the negative mass shell is determined by the {\it dual} representation of the deformed Lie $\kappa$-Poincar\'e algebra, which, as discussed in more detail below, is defined through the antipode map of the generators. As we will see this will play a crucial role in our description of the charge conjugation operator.

In preparation for our discussion of discrete symmetries for $AN(3)$-valued momenta let us recall some basic facts about the $\kappa$-Poincar\'e algebra. With a choice of translation generators given by the `embedding coordinates' defined by (\ref{1}) the Lorentz algebra is not modified and reads
\begin{align}
[N_i, P_j] &= i\delta_{ij}\, P_0\,,\quad [N_i, P_0] =i P_i\nonumber\\
[M_i, M_j] &= i\epsilon_{ijk} M_k\,,\quad [M_i, N_j] = i\epsilon_{ijk} N_k\,,\quad [N_i, N_j] =- i\epsilon_{ijk} M_k\label{4}\,.
\end{align}
All the non-trivial features due to the group structure of momentum space appear in the way one defines dual and tensor product representations of the algebra \cite{Arzano:2014cya}, as we discuss below. Let us notice here that the algebra structure (\ref{4}) is invariant under an arbitrary linear transformation of the generators $T_i\rightarrow A_i^j\, T_j$, in particular under transformations of the form $T_i\rightarrow - T_i$ for some or all generators. If there is no scale available such linear transformations exhaust all the possible isomorphism of the algebra. If, however, a scale is available, as is the case when momentum space has curvature, we can replace all the generators by the antipodal map.

In order to introduce such structure let us recall (the known fact) that the states of an ordinary massive scalar relativistic particle in Minkowski space belong to unitary irreducible representations of the Poincar\'e group. These can be identified with smooth functions on the positive mass-shell and, in the usual textbook treatment, such states are denoted by kets $| k \rangle$, being eigenstates of translation generators $P_{\mu} | k \rangle = k_{\mu}| k \rangle$. The action of such generators on dual states and on tensor product states is given, respectively by the {\it dual} and the {\it tensor product} representation of the algebra of translation generators: $P_{\mu} \langle k | = - k_{\mu} \langle k | = \langle k | (-P_{\mu}) \equiv \langle k | S(P_{\mu} )$ and $P_{\mu} (| k_1 \rangle \otimes | k_2 \rangle) =  P_{\mu} | k_1 \rangle \otimes | k_2 \rangle +| k_1 \rangle \otimes P_{\mu} | k_2 \rangle \equiv \Delta P_{\mu} | k_1 \rangle \otimes | k_2 \rangle$. We introduced the {\it antipode} $S(P_{\mu} )= - P_{\mu}$ and {\it co-product} $ \Delta P_{\mu}= P_{\mu} \otimes \mathbbm{1} + \mathbbm{1} \otimes P_{\mu}$, borrowing notation and terminology from the theory of Hopf algebras.

When the linear space of the momentum four-vectors is replaced by  the group $AN(3)$ we can define one-particle states in a straightforward fashion: as in the ordinary Minkowski space case, we denote such states with kets $|h \rangle$ labelled by group elements belonging to a given orbit of $SO(3,1)$ on $AN(3)$.
For any parametrization of momentum space we can define a set of four-momentum observables $P_{\mu}$ which associate four-vectors to eigen-kets $P_{\mu} |h \rangle = p_{\mu} (h) |h\rangle$. For the action on dual states the non-trivial structure of momentum space comes into play and one has that
\be
P_{\mu}\, \langle h | = p_{\mu}(h^{-1})\langle h | \equiv \langle h | S(P_{\mu})\,,
\ee
indicating that the eigenvector associated to such states `reads off' the coordinates of the {\it inverse} momentum $h^{-1}$ and in general $p_{\mu}(h^{-1})\neq -p_{\mu} (h)$. Likewise for tensor product states $P_{\mu} (| h_1 \rangle \otimes | h_2 \rangle) = p_{\mu}(h_1 h_2) | h_1 \rangle \otimes | h_2 \rangle \equiv \Delta P_{\mu} | h_1 \rangle \otimes | h_2 \rangle$. The non-abelian group multiplication is reflected in a {\it non-trivial co-product} for $P_{\mu}$ so that the total four-momentum of the composite state $| h_1 \rangle \otimes | h_2 \rangle$ is given by the coordinates of the product group element $h_1 h_2$ and implies a {\it non-abelian composition law}
\begin{equation}\label{nabcomp}
p_{\mu}(h_1 h_2) = p_{\mu}(h_1) \oplus p_{\mu}(h_2)\,,	
\end{equation}
which in our case takes the form
\begin{align}
\left(p^{(1)}\oplus p^{(2)}\right)_0 &=\frac{1}\kappa\, p^{(1)}_0\left(p_0^{(2)}+p_4^{(2)}\right)+\kappa\, p^{(2)}_0\left(p_0^{(1)}+p_4^{(1)}\right)^{-1}+\left(p_0^{(1)}+p_4^{(1)}\right)^{-1}\mathbf{p}^{(1)}\cdot\mathbf{p}^{(2)}\label{sum0}\\
\left(p^{(1)}\oplus p^{(2)}\right)_i &=\frac{1}\kappa\, p_i^{(1)}\left(p_0^{(2)}+p_4^{(2)}\right)+p_i^{(2)}\label{sumi}
\end{align}
Let us stress that this momentum composition rule implies that the ``inverse'' momentum is defined by $p_\mu(h^{-1})$, which is the eigenvalue of the operator $S(P_\mu)$. In what follows, rather than to the unusual composition rule (\ref{nabcomp}), we will turn our attention to the antipodes of the generators of the $\kappa$-Poincar\'e algebra, since they will play a key role in the definition of discrete symmetries. In particular for translation generators associated to the embedding coordinates
defined by (\ref{1}) the antipodal map for momenta takes the form
\begin{equation}\label{5}
    S(P)_0  =-P_0+\frac{\mathbf{P}^2}{P_0+P_4}=-P_4+\frac{\kappa^2}{P_0+P_4}, \quad S(P)_i  =-\frac{\kappa {P}_i}{P_0+P_4}\, .
\end{equation}
Notice that in the limit $\kappa\rightarrow\infty$, $S(p)_0=-P_0$ and $S(p)_i=-P_i$. Also $S(P_0)<0$ for $P_0$ positive, and $S(P)_0^2 - S(P)_i^2 = P_0^2-P_i^2$ so the antipode mapping preserves the mass shell condition (it maps a positive energy state into the negative energy one of the same mass.)

Inspecting eqs.\ (\ref{5}) we notice that the antipodes of momenta are bounded from above, for example
\begin{equation}\label{10a}
    -S( P)_0=P_4-\frac{\kappa^2}{P_0+P_4}\doteq\sqrt{\kappa^2+M^2} - \frac{\kappa^2}{P_0+\sqrt{\kappa^2+M^2}}<\sqrt{\kappa^2+M^2}
\end{equation}
This seems to be very puzzling at the first sight, because it follows from (\ref{10a}) that the energy of antiparticles is bounded. However one should remember that the deformed particle kinematics we are discussing here is intended as a possible approximation of a full quantum gravity theory at energies below the Planck scale, and thus features of the deformed theory should be taken seriously below the onset of full quantum gravity regime, i.e., far below the Planck scale. For this reason it is convenient to write down the leading order contributions to eq.\ (\ref{5}), which are useful for phenomenological analyses, to wit
\begin{equation}\label{5a}
  S(P)_0  =-P_0+\frac{\mathbf{P}^2}{\kappa} + O\left(\frac1{\kappa^2}\right), \quad S(P)_i  =-P_i +\frac{P_0 {P}_i}{\kappa}+ O\left(\frac1{\kappa^2}\right)\, .
\end{equation}
For Lorentz generators the antipodes are given by
\begin{equation}\label{6}
  S(M)_i = - M_i\,,\quad S(N)_i =-\frac{1}\kappa\,(P_0+P_4) N_i +\frac{1}\kappa\, \epsilon_{ijk}\, P_j M_k\, ,
\end{equation}
with the leading order expansion
\begin{equation}\label{6a}
  S(N)_i=-N_i+\frac{1}\kappa\,\left(-P_0 N_i + \epsilon_{ijk}\, P_j M_k \right)+ O\left(\frac1{\kappa^2}\right)\,.
\end{equation}
Since the $\kappa$-Poincar\'e algebra is a Hopf algebra, the antipode satisfies the Hopf algebra-defining axioms making it consistent with the algebra and co-algebra structure. First, it can be verified by direct calculation that the antipodes (\ref{5}) and (\ref{6})  map the Lorentz algebra into itself, exactly as it was in the classical case
\begin{align}
[S(N)_i, S(P)_j] &=- i\delta_{ij}\, S(P)_0\,,\quad [S(N)_i, S(P)_0] = -iS(P)_i \,,\quad [S(M)_i, S(M)_j] =- i\epsilon_{ijk} S(M)_k \nonumber\\
& [S(M)_i, S(N)_j] =- i\epsilon_{ijk} S(N)_k\,,\quad [S(N)_i, S(N)_j] =  i\epsilon_{ijk} S(M)_k \,. \label{7}
\end{align}
As a consequence of this the Casimir is invariant under the operation of taking antipodes. 

Further, let us notice that the Hopf algebraic consistency condition between the antipode and coproduct
$$
\cdot(S\otimes {\rm id})\circ \Delta =\cdot({\rm id}\otimes S)\circ \Delta =0
$$
has a simple physical interpretation in the case of momenta. Namely if $|h\rangle$ is a one particle state with momentum $P_\mu|h\rangle = p_\mu(h) |h\rangle$, then the eigenvalue of $S(P)_\mu$, given by $S(P)_\mu|h\rangle = \ominus p_\mu(h) |h\rangle \equiv p_\mu(h^{-1}) |h\rangle $, is the inverse momentum, i.e. $(\ominus p)\oplus p = p\oplus (\ominus p)=0$.
It can be also checked by explicit calculation that, as in the undeformed case, for $P_\mu$ and $M_i$ the composition of two antipode operations (\ref{5}), (\ref{6}) is an identity $S^2=1$. This property is obvious for rotations, since $M_i$ are undeformed and not surprising for group valued momenta for which the momentum obtained by the action of the antipode is given by the inverse group element $\ominus p_\mu(h) = p_\mu(h^{-1})$, and it follows that $S^2=1$. The situation is different for boosts, however. Indeed using the anti-homomorphism property of the antipode, $S(ab)=S(b)S(a)$, one finds that
\begin{equation}\label{SN}
 S^2(N_i) = -\frac1\kappa\, S(N_i)\,S(P_0+P_4) + \frac1\kappa\,\epsilon_{ijk}\,S(M_k)\, S(P_j) = N_i +\frac{3}\kappa\, S(P_i)\,,
\end{equation}
where $S(P_i)$ is given by (\ref{5}). This is a particular case of the general formula quoted in \cite{Borowiec:2013lca}.

%The fact that $S^2 \neq 1$ for boost might have interesting consequences. In particular it is known that for the Rindler observer the energy is defined by boost hamiltonian, and discrete symmetries transform one Rindler wedge into another. Therefore, the nontrivial form of (\ref{SN}) presumably influences the physics of Unruh effect, and as a consequence, of Hawking radiation. We will investigate this circle of questions in a separate paper.

We can now turn to the discussion of discrete transformations. Our strategy will be to focus on the physical states of the quantum system and derive discrete transformations as symmetries involving the quantum charges carried by such states.\footnote{After completing this note we learned from G. Amelino-Camelia that the idea that deformed discrete symmetries might require the concept of antipode for their description appeared a while ago, in a preliminary form, in \cite{AmelinoCamelia:1999pm}.}

\section{Space and time reflection}
Our starting point will be space reflection or {\it parity}. This symmetry can be discussed using the simplest quantum states, those of a scalar neutral particle, which can be fully characterized by the (on-shell) energy and momentum they carry. The one-particle Hilbert space $\mathcal{H}$ can be viewed as a unitary irreducible representation of the Poincar\'e algebra (like for an undeformed particle) \cite{Ruegg:1994bk} and it is spanned by kets given by $|k\rangle \equiv |\omega_{\bf{P}}, \bf{P}\rangle$, with $\omega_{\bf{P}}=\sqrt{{\bf{P}}^2+M^2}$. Given the striking similarity with the undeformed case one might be tempted to stick to the ordinary definition of space inversion which leaves unaffected the particle's energy $\omega_{\bf{P}}$ while inverting its spatial momentum ${\bf P} \rightarrow -{\bf P}$.
%At a first inspection it might seem like one has two possible choices: stick with the usual description of parity transformation ${P}_i \rightarrow -{P}_i$ or make use of the antipode map $P_i \rightarrow S(P)_i$.
There is, however, a crucial property of the usual formulation of particle kinematics that must still hold in the deformed setting. Namely that the total linear momentum of a particle and of its parity image must vanish. This must hold true even when their momentum belong to a non-abelian Lie group as is the case of deformations we are considering. Indeed a non-vanishing momentum will mean that the pair moves in a specific direction thus violating the very notion of parity symmetry. We are thus led to adopt the antipode in our definition of parity transformed translation generators $\mathbb{P}:{P}_i \rightarrow S(P)_i$. An important point is that if we want to use the antipode we must use it {\em for all the generators} in order to preserve the form of the invariants, like the mass Casimir, and the form of the algebra. Thus we are led to define the following action for the parity operator $\mathbb{P}$ on $\kappa$-Poincar\'e generators
\begin{align}
   \mathbb{P}({P}_i)  = S(P)_i, \quad \mathbb{P}(P_0) =-S (P)_0\nonumber\\
   \mathbb{P}({M}_i) =- S(M)_i, \quad \mathbb{P}(N_i) =S (N)_i\label{8}
\end{align}
The action of $\mathbb{P}$ on rotation generators is obtained from the requirement that in the undeformed limit parity does not change spin and thus does not change the sign of $M_i$. Similarly for the action of $\mathbb{P}$ on boosts, which in the undeformed limit switches their signs. Notice the ordinary behaviour of boosts under parity can be easily understood if one recalls that, in the co-adjoint orbit picture of a particle phase space \cite{Kir}, boost generators can be used to parametrize space coordinates and thus must change sign under parity. Thus parity can be fully characterized in a purely algebraic fashion using the generators of the (deformed) Poincar\'e algebra.

The same holds true for {\it time inversion}. Such symmetry is again a map of one-particle states into themselves and again we can start by the physical requirement that time reflection $\mathbb{T}$ is such that the particle and its time-reflected image have total vanishing momentum. Thus we must have again the generator of spatial translation to be mapped into its antipode $\mathbb{T}({P}_i) = S(P)_i$ and all the other transformations must involve the antipodes. As in the case of parity we can derive the action of $\mathbb{T}$ on the other generators by requiring that its net effect will be to map the algebra \eqref{4} into \eqref{7} and that in the $\kappa\rightarrow \infty$ we recover the usual action of the time reversal operator. Thus in such limit rotation generators should flip sign since time reversal inverts spin and boost and time translation generators should remain unaffected.
% Is there a physical reasoning similar to the one we followed for parity which allows us to determine the action of a time reversal operator  on Lorentz generators? We can require that time reflection again is such that the particle and the time reflected image have total vanishing momentum. The other requirement for such discrete symmetry will be that they also have vanishing spin, since time reversal flips spins, and thus $\mathbb{T}\, {M}_i \, \mathbb{T}^{-1} = S(M)_i$ so
 Thus the full set of time reflected generators is given by
\begin{align}
   \mathbb{T}({P}_i) = S(P)_i, \quad \mathbb{T}(P_0) =-S (P)_0\nonumber\\
   \mathbb{T}({M}_i) = S(M)_i, \quad \mathbb{T}(N_i) =-S (N)_i\,.\label{9}
\end{align}
Notice that it follows from (\ref{8}) and (\ref{9}) that the combination of deformed parity and time reversal $\mathbb{P}\mathbb{T}$ is actually {\it undeformed} for momenta and rotations, as a result of the fact that  for them the antipode map is an involution, i.e. $S^2=1$, but it becomes nontrivial for boost, as a consequence of (\ref{SN}).

\section{Charge conjugation}
%{\bf (4)}: Hilbert space complex scalar field, momentum and charge labels.\\
In order to understand charge conjugation we need to consider a system which possesses a `charge' quantum number associated to a complex structure defined on its Hilbert space. This calls for a brief detour to review how such Hilbert space is constructed from the classical system. The phase space of a classical real scalar field can be equivalently described in terms of the space of solutions of the Klein-Gordon equation, in coordinate space, or by complex functions on the mass-shell, in momentum space. Both spaces carry a representation of the Poincar\'e group. The Hilbert space describing the states of the associated quantum field can be obtained constructing a unitary irreducible representation out of the classical phase space. This is achieved via the introduction of a complex structure which amounts to a choice of time direction or {\it positive energy} \cite{Geroch:2013i} . Put it plainly such structure dictates the way we multiply a function on the mass shell by a complex number. For functions on the positive mass shell we multiply by the complex number while for functions on the negative mass shell we multiply by its complex conjugate (this ensures that, for a complex field, antiparticle states have positive energy). Thus, if functions on the positive mass-shell, appropriately equipped with a positive definite inner product, give us the one-particle Hilbert space $\mathcal{H}$, functions on the negative mass shell are naturally associated to the complex conjugate space $\bar{\mathcal{H}}$. It is a known fact that $\bar{\mathcal{H}}$ is isomorphic to the dual Hilbert space $\mathcal{H}^*$ and thus, if we denote states in $\mathcal{H}$ by kets $|k\rangle$, the bras $\langle k|$ are naturally seen as elements of $\mathcal{H}^*\simeq \bar{\mathcal{H}}$. If we characterize such states in terms of their quantum charges associated to translation generators, given the definition of dual representation of a Lie algebra, one can identify $\langle -k| \equiv |k\rangle$ which is nothing but a reflection of the familiar reality condition for the field $\bar\phi(-k) = \phi(k)$ which provides a natural isomorphism between $\mathcal{H}$ and $\bar{\mathcal{H}}$.

For complex fields the Hilbert spaces $\mathcal{H}$ and $\bar{\mathcal{H}}$ can no longer be identified via the reality condition and, in fact, they represent the one-particle and {\it one-antiparticle} Hilbert spaces respectively. Now the map $\mathbb{C}: \phi(k)\rightarrow \bar\phi(-k)$ relates such two spaces and it corresponds to the ordinary {\it charge conjugation}, a map swapping particles with antiparticles. How can this picture be extended to states in which momentum labels are coordinates on a group manifold?
%{\bf (1)}
The key is the isomorphism between $\bar{\mathcal{H}}$ and $\mathcal{H}^*$ which tells us that once again our map involves the antipode. Indeed charge conjugation will again map states in $\mathcal{H}$ to states in $\bar{\mathcal{H}}$ and according to the isomorphism above this will map a representation of the Lorentz generators into its dual and thus all generators will have to be mapped once again into their antipodes.
%This can be seen as follows \cite{Arzano:2009ci}. Since the space-time associated with a momentum space of non-trivial geometry is necessarily associated with  non-commutative spacetime, the complex conjugation of plane waves is non-trivial. If one assumes that the product of the plane wave and its conjugate is to be equal 1, then in the case of the AN(3) plane waves one easily concludes that the conjugation replaces all the momenta $k$ with their antipodes.\\
To find the correct form of the map, however, we must keep in mind that, as explained above, antiparticle states are multiplied by the complex conjugate of a complex number. Thus the deformed charge conjugation operator will have again to be such that the algebra \eqref{4} is mapped into \eqref{7} but now all the generators in the $\kappa\rightarrow \infty$ limit should remain unchanged.  This fixes the transformation to the following form
\begin{align}
   \mathbb{C}({P}_i) = - S(P)_i, \quad \mathbb{C}(P_0) =-S (P)_0\nonumber\\
   \mathbb{C}({M}_i)  =  - S(M)_i, \quad \mathbb{C}(N_i) = -S (N)_i\,.
\end{align}
%we need to consider a quantum relativistic system which exhibits anti-particles. The simplest such system is a complex scalar field. \\
%One-particle Hilbert space $\mathcal{H}\equiv C^{\infty}(M^+_m)$. One-antiparticle Hilbert space given by {\it complex conjugate} space  $\bar{\mathcal{H}}\equiv C^{\infty}(M^-_m)$. Such space is isomorphic to the {\it dual} Hilbert space $\mathcal{H}^*$ of functions from $\mathcal{H}$ to the complex numbers and thus to each element $\bar{|k\rangle}\in \bar{\mathcal{H}}$ we can associate a bra $\langle k'| \in \mathcal{H}^*$. Most importantly, such correspondence leads to an isomorphism $\bar{\mathcal{H}}^*\simeq \mathcal{H}$, which, in the case of a complex field, is nothing but {\it charge conjugation}. In terms of functions on the  mass shell such map will be given by [Geroch]
%\begin{equation}
%C: \phi(k)\in C^{\infty}(M^+_m) \longrightarrow \bar{\phi}(- k)\in C^{\infty}(M^-_m)\,.
%\end{equation}
Such map exchanges the role of {\it deformed} particle and anti-particle and it can be easily showed that it exchanges the roles of the deformed $U(1)$ charges carried by particle and antiparticle states derived in \cite{Arzano:2009ci} for a complex scalar field. Notice also that since the antipode preserves the mass-shell relation, particles and antiparticles have the same masses, as in ordinary QFT.
 %\cite{Arzano:2009ci} that  $\mathbb{C}$ defined in above is an operator relating the positive and negative $U(1)$ charge modes of the complex scalar field.

%We thus write for $\mathbb{C}$ the following relations
%\begin{align}
   %\mathbb{C}\, {P}_i \, \mathbb{C}^{-1} = - S(P)_i, \quad \mathbb{C}\, P_0 \, \mathbb{C}^{-1} =-S (P)_0\nonumber\\
   %\mathbb{C}\, {M}_i \, \mathbb{C}^{-1} = - S(M)_i, \quad \mathbb{C}\, N_i \, \mathbb{C}^{-1} =-S (N)_i\label{9}
%\end{align}

%\section{CPT}

\section{Summary}
We provided the first formulation of discrete symmetries in a model of deformed relativistic symmetries defined on a group manifold momentum space. We focused, in particular, on the $\kappa$-Poincar\'e algebra, a widely studied example of deformed relativistic kinematics based on de Sitter momentum space. Even though for definiteness we restricted to such particular model, it should be noted that our construction is based on general features of relativistic kinematics with group valued momenta and as such it can be applied to other, structurally similar, frameworks.

A key ingredient of our construction was the use of the ``classical basis" of the $\kappa$-Poincar\'e algebra whose algebraic sector (\ref{4}) is described by an ordinary Lie algebra. This makes it possible for us to use the ordinary minus "-" in the definition of the discrete transformation action. In other, nonlinear bases this minus will have to be replaced by a nontrivial, nonlinear operation. We will address the question of the form of discrete symmetries in other $\kappa$-Poincar\'e bases in a forthcoming work.

Our main aim was to provide a starting point and the basic tools for an analysis of possible phenomenological signatures of $\kappa$-deformations and similar models in high sensitivity tests of discrete symmetries and of CPT invariance.  As it is clear from the construction above the action of the $\mathbb{CPT}$ operator is `deformed', albeit in a subtle way. Indeed it follows from the discussion above that
\begin{align}
   \mathbb{CPT}({P}_i) = - S(P)_i, \quad \mathbb{CPT}(P_0) =-S (P)_0\nonumber\\
   \mathbb{CPT}({M}_i)  = S(M)_i, \quad \mathbb{CPT}(N_i) = S (N)_i+\frac{3}{\kappa}\, P_i\,.\label{10}
\end{align}
Notice that such departure from the ordinary CPT map is {\it universal}, it is related to a non-trivial geometry of momentum space and as such it affects all matter and interactions. The next step beyond the basic analysis we reported here, will be to provide a systematic investigation of the unconventional behaviour of all three discrete symmetries and their combinations, in various phenomenological frameworks which can be sensitive to the corrections introduced by the (Planckian) curvature of momentum space. We defer this to upcoming works.

\section*{Acknowledgements}

We thank G.\ Amelino-Camelia, A.\ Borowiec, A.\ Di Domenico, and G. Gubitosi for useful discussions. JKG acknowledges support by the Polish National Science Center under the agreements DEC- 2011/02/A/ST2/00294, 2014/13/B/ST2/04043.

\end{document}